\documentclass[11pt]{article}
\usepackage[a4paper,bindingoffset=0.2in,%
             left=0.9in,right=0.9in,top=1in,bottom=1in,%
             footskip=.25in]{geometry}
 
\usepackage{amsmath,amssymb,amsthm,bm,mathtools}
\usepackage{graphicx}
\usepackage{booktabs}
\usepackage{microtype}
\usepackage{hyperref}
\usepackage{authblk}
\allowdisplaybreaks

\newcommand{\mbf}{\mathbf}

\newcommand{\mcl}{\mathcal}
\newcommand{\mrm}{\mathrm} 
\usepackage{comment}
\usepackage{xcolor}
\usepackage{xstring}

\def\C{\mathbb C}

\newcommand*{\norm}[1]{\left\|#1\right\|}

\newcommand*{\card}[1]{\left|#1\right|}

\title{Low-Dimensional Hermite-Gaussian Representations for Ptychographic Reconstruction}
\author[1]{Arya Bangun}
\author[2]{Evan Laksono}

\affil{Forschungszentrum Jülich,  $^2$Stanford University}

\date{}

\begin{document}
\maketitle
\begin{abstract}
 
We analyze the reconstruction error of a low-dimensional Hermite--Gaussian (HG) ptychographic method used in real-time Wigner distribution deconvolution (WDD) reconstruction. We derive an upper bound on the discrepancy between the low-dimensional, aberration-corrected reconstruction and classical Fourier-domain WDD. The bound decomposes the reconstruction error into contributions from HG--Fourier basis mismatch and aberration-coefficient mismatch, explicitly characterizing their dependence on the number of retained HG modes, Wiener regularization, aberration mismatch, and measurement dose. Numerical experiments on simulated electron and X-ray ptychographic data characterize these dependencies and demonstrate robustness to Poisson noise. These results provide theoretical and  guidance for selecting the reduced representation and reconstruction parameters, enabling computationally and memory-efficient for near-real-time ptychographic reconstruction.
\end{abstract}
\section{Introduction}

Ptychography is a computational imaging technique that reconstructs a complex-valued object from a collection of diffraction measurements acquired at overlapping illumination regions
\cite{Hoppe1969, Hoppe1969a, Hoppe1969b, rodenburg1992theory, maiden2009improved}.
By exploiting the redundancy between neighboring probe positions, ptychography enables the recovery of both amplitude and phase information from intensity-only diffraction measurements.
It has become an important imaging modality across electron, X-ray, and optical microscopy, providing high spatial resolution, quantitative phase contrast, and robustness to imperfections in the imaging system
\cite{thibault2008high, zheng2013wide, jiang2018electron}.

In scanning electron and X-ray microscopy, a two-dimensional diffraction pattern is recorded at each two-dimensional probe position, resulting in four-dimensional scanning diffraction datasets
\cite{ophus2019four, thibault2008high, favre2020pynx}.
These measurements contain rich information about the specimen, the illuminating probe, and experimental parameters such as optical aberrations.
Recovering this information, however, generally requires solving a large-scale inverse problem.
In practical experiments, the reconstruction may involve not only the specimen transmission function or projected electrostatic potential, but also unknown probe parameters, aberration coefficients, partial coherence, scan distortions, and measurement noise.
Many existing approaches rely on iterative optimization, which can be computationally demanding and sensitive to initialization, regularization, and model mismatch
\cite{jiang2018electron, bangun2022inverse, chen2021electron, schropp2020ptynami}.

The computational challenge is further amplified by the dimensionality of the measurements.
A two-dimensional diffraction image is acquired at every two-dimensional scan position, producing datasets indexed simultaneously by real-space scan coordinates and reciprocal-space detector coordinates.
As detector frame rates, detector sizes, and scan dimensions continue to increase, the resulting datasets can become prohibitively large for conventional reconstruction pipelines
\cite{spurgeon2021towards, batey2022high, plotkin2022100}.
This motivates the development of low-dimensional representations that preserve the information relevant to ptychographic reconstruction while reducing memory requirements and computational cost, particularly for live or near-real-time reconstruction during experiments
\cite{strauch2021live, bangun2023wigner, pelz2021real, welker2024live}.
Although recent work has demonstrated the potential of reduced-dimensional representations for accelerating ptychographic reconstruction, a quantitative characterization of the reconstruction error introduced by such representations, particularly in the presence of aberration mismatch, regularization, and measurement noise, remains limited.

In this work, we analyze a low-dimensional ptychographic reconstruction framework based on Hermite--Gaussian (HG) representations of the diffraction data.
The HG basis is particularly well suited to diffraction imaging because Hermite--Gaussian functions form a complete orthonormal eigenbasis of the Fourier transform, allowing Fourier-domain operations to be represented naturally in the same modal basis.
By retaining only a subset of HG modes, the detector-space data can be represented in a reduced form, enabling a low-dimensional formulation of Wigner distribution deconvolution (WDD) and substantially decreasing the dimensionality of the reconstruction problem. We derive an upper bound on the discrepancy between the low-dimensional, aberration-corrected reconstruction and the classical Fourier-domain WDD reconstruction.
Numerical experiments with simulated electron and X-ray ptychographic data validate these dependencies, identify the corresponding low-error operating regimes, and demonstrate the effects of Poisson noise and measurement dose on reconstruction accuracy.
\section{Definition and Background}

In this section, we briefly introduce the notation used throughout this work as well as the problem formulation.

\paragraph{Notations.}
Vectors are written in lowercase bold letters, e.g., $\mathbf r$. For a complex variable $c$, the complex conjugate is denoted by $\overline{c}$. We use the following Fourier convention:
\begin{equation*}
\mathcal F_{\mathbf r\to\mathbf q}\{f\}(\mathbf q)
=
\int_{\mathbb R^2}
f(\mathbf r)
e^{-2\pi i\mathbf q\cdot\mathbf r}
\,d^2r,
\qquad
\mathcal F^{-1}_{\mathbf q\to\mathbf r}\{F\}(\mathbf r)
=
\int_{\mathbb R^2}
F(\mathbf q)
e^{2\pi i\mathbf q\cdot\mathbf r}
\,d^2q.
\end{equation*}
The Hermite--Gaussian (HG) basis function is given by
\cite{thangavelu1993lectures, teschl2014mathematical}
\begin{equation*}
\phi_n(\xi)
=
\frac{1}{\pi^{1/4}\sqrt{2^n n!}}\,
H_n(\xi)\,
e^{-\xi^2/2},
\end{equation*}
where
\begin{equation*}
H_n(\xi)
=
(-1)^n e^{\xi^2}
\frac{d^n}{d\xi^n}
e^{-\xi^2}
\end{equation*}
is the $n-$th Hermite polynomial. The two-dimensional separable Hermite--Gaussian basis is
\begin{equation*}
\Phi_{\mathbf n}(\mathbf q)
=
\phi_{n_x}(\xi_x)\,
\phi_{n_y}(\xi_y),
\qquad
\mathbf n=(n_x,n_y).
\end{equation*}
Let
$\{\Phi_{\mathbf n}(\mathbf q)\}_{\mathbf n\in\mathbb N^2}$
be an $L^2(\Omega_q)$-orthonormal HG family such that
\begin{equation*}
\int_{\Omega_q}
\Phi_{\mathbf n}(\mathbf q)
\overline{\Phi_{\mathbf m}(\mathbf q)}
\,d^2q
=
\delta_{\mathbf n\mathbf m}.
\end{equation*}
The HG coefficient transform is defined as
\begin{equation}
\mathcal{HG}_{\mathbf q\to\mathbf n}
\{F(\mathbf q )\}
:=
\int_{\Omega_q}
F(\mathbf q)
\overline{\Phi_{\mathbf n}(\mathbf q)}
\,d^2q.
\end{equation}
Throughout this article, we write the Wiener map as
\begin{equation*}
W_\epsilon(z)
:=
\frac{\overline z}{|z|^2+\epsilon},
\end{equation*}
where $\epsilon>0$. We use the standard bounds
\begin{equation*}
|W_\epsilon(z)|
\le
\frac{1}{2\sqrt{\epsilon}},
\qquad
|W_\epsilon(z_1)-W_\epsilon(z_2)|
\le
\frac{1}{\epsilon}|z_1-z_2|,
\end{equation*}
see, e.g.,
\cite{hofmann2006approximate, engl1996regularization}. The $ L^p-$norm of a function $ f $ over a domain $ \Omega  $ is defined as $\|f\|_{L^p(\Omega)}
=
\left(
\int_{\Omega} |f(x)|^p\,dx
\right)^{1/p}
\quad \text{for} \quad 1\leq p<\infty
 $ and $\|f\|_{L^\infty(\Omega)}
=
\operatorname*{ess\,sup}_{x\in\Omega}|f(x)| \quad \text{for} \quad   p=\infty.$

\paragraph{Ptychography Model.}
Suppose that a focused electron probe $P(\mbf r)$ illuminates a specimen object $O(\mbf r)$, where $\mbf r=(x,y)$ is the transverse spatial coordinate. The forward model of 4D-STEM data for a relatively thin specimen is given by
\begin{equation}
I(\mbf q,\mbf R)
=
\card{
\mcl F_{\mbf r\to\mbf q}
\left(
O(\mbf r)P(\mbf r-\mbf R)
\right)
}^2.
\label{eq:forward_model}
\end{equation}
Here, the focused probe is shifted over the scan coordinate
$\mbf R=(R_x,R_y)$, and the far-field detector records the diffraction intensity
$I(\mbf q,\mbf R)$ in reciprocal space.

\begin{figure}[!htb]
    \centering
    \includegraphics[width=1.\linewidth]{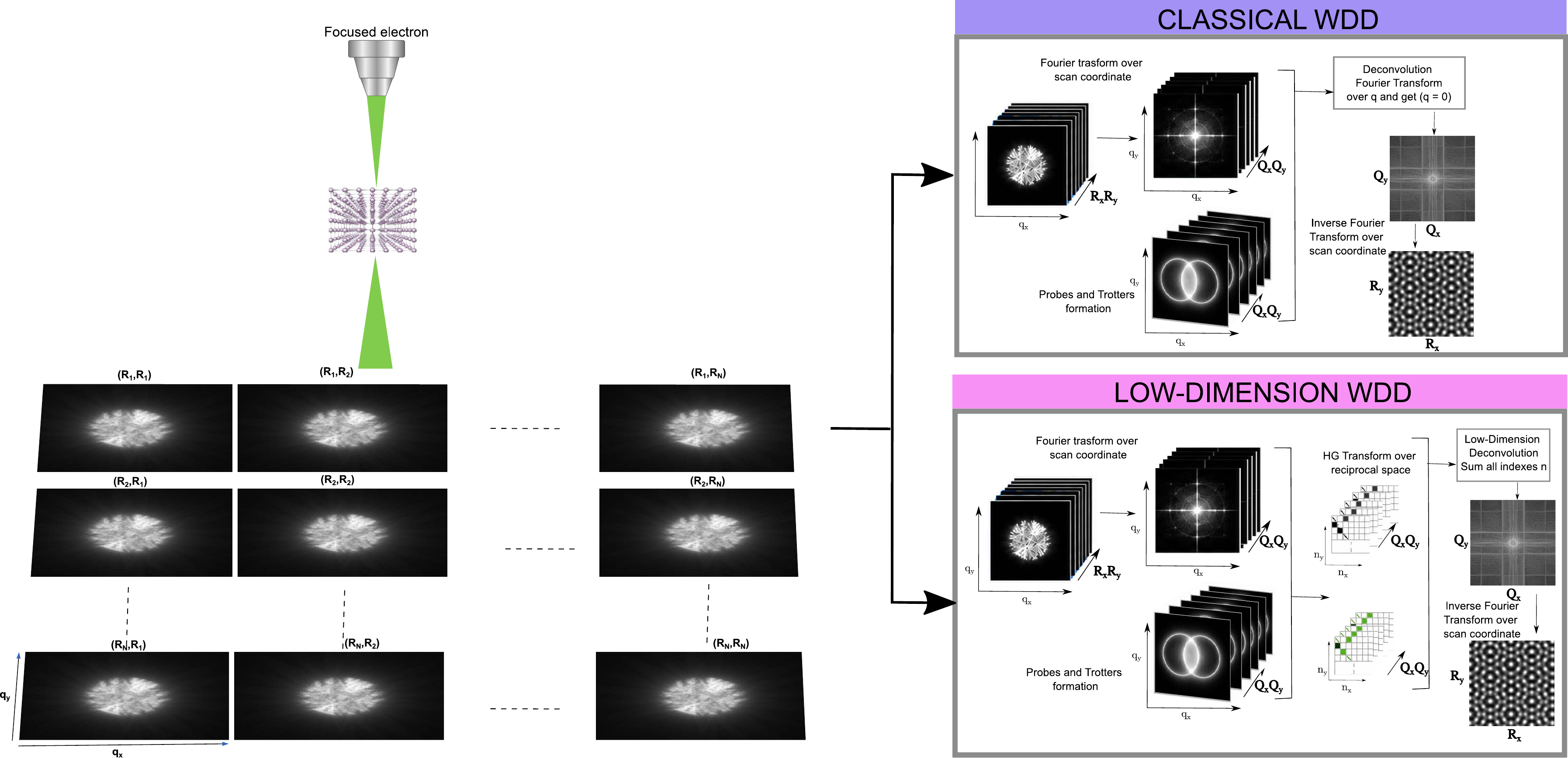}
    \caption{
    Visualization of 4D-STEM acquisition. The focused electron probe is rastered over scan positions $\mbf R$, and a two-dimensional diffraction intensity is recorded in reciprocal-space coordinate $\mbf q$ at each position. The proposed approach represents the detector-space information in a low-dimensional HG basis while preserving the structure required for WDD reconstruction.
    }
    \label{fig:stem_model}
\end{figure}

To model realistic measurement data arising from electron counting at the detector, we assume that the measured intensity follows Poisson statistics. In particular,
\begin{equation}
I^{\mrm{meas}}(\mbf q,\mbf R)
\sim
\operatorname{Poisson}
\left(
\eta\, I(\mbf q,\mbf R)
\right),
\label{eq:poisson_model}
\end{equation}
where $\eta>0$ is a scaling factor chosen according to the prescribed electron dose. A larger value of $\eta$ therefore corresponds to a larger expected number of detected electrons. The detected electron count associated with a scan position $\mbf R$ is denoted by
\begin{equation}
D(\mbf R)
=
\int_{\Omega_q}
I^{\mrm{meas}}(\mbf q,\mbf R)\,d^2q,
\end{equation}
and we define
\begin{equation}
D_{\max}
=
\sup_{\mbf R}D(\mbf R),
\qquad
D_{\rm tot}
=
\int_{\Omega_R}
D(\mbf R)\,d^2R.
\end{equation}
The same statistical model applies to X-ray ptychography, where the measured signal is governed by photon-counting statistics. In this case, the corresponding dose represents the expected number of detected photons rather than electrons, while the Poisson measurement model and subsequent reconstruction analysis remain unchanged. The acquisition geometry is illustrated in Fig.~\ref{fig:stem_model}.
The object contains the projected electrostatic potential $V(\mbf r)$ through
\begin{equation*}
O(\mbf r)
=
e^{i\sigma V(\mbf r)},
\end{equation*}
where $\sigma$ is the interaction parameter. The probe is modeled as
\begin{equation}
P(\mbf r)
=
\mcl F_{\mbf q\to\mbf r}^{-1}
\left(
\card{A(\mbf q)}e^{i\chi(\mbf q)}
\right),
\label{eq:probe_recip}
\end{equation}
where $A(\mbf q)$ is the aperture function and $\chi(\mbf q)$ is the aberration function. The aberration phase is expanded as
\begin{equation}
\chi(\mathbf q)
=
\frac{2\pi}{\lambda}
\sum_{l=1}^{L}
c_l\,\Psi_l(\mathbf q),
\end{equation}
where $c_l$ and $\Psi_l$ denote the coefficient and basis function of the $l-$th aberration mode, respectively, including defocus, astigmatism, coma, trefoil, and spherical aberration
\cite{krivanek1999towards, haider1998electron}. Vectorizing the aberration coefficient, we write the notation as $\mbf c \in \C^L.$ 

The single index $l$ is used to flatten the Krivanek mode labels, i.e.,
\begin{equation*}
l\leftrightarrow(n,m,s),
\qquad
s\in\{a,b\},
\end{equation*}
so that
$c_l=c_{nm,s}$ and
$\Psi_l=\Psi_{nm,s}$.
Following the Krivanek notation, the aberration basis functions are written as
\begin{equation}
\Psi_{nm,a}(q,\phi)
=
\frac{(\lambda q)^{n+1}}{n+1}\cos(m\phi),
\qquad
\Psi_{nm,b}(q,\phi)
=
\frac{(\lambda q)^{n+1}}{n+1}\sin(m\phi),
\label{eq:krivanek}
\end{equation}
where $q=|\mathbf q|$ and $\phi$ is the azimuthal angle of $\mathbf q$.

For the noiseless forward model, expanding the squared magnitude and applying the Fourier transform over the scan coordinate
$\mathcal{F}_{\mathbf R\to\mathbf Q}$ yields the WDD convolution form. We denote
\begin{equation*}
I(\mbf q,\mbf Q)
=
\mathcal{F}_{\mathbf R\to\mathbf Q}
\left\{
I(\mbf q,\mbf R)
\right\},
\end{equation*}
such that
\begin{align*}
I(\mbf q,\mbf Q)
&=
O(\mbf q)
\overline{O(\mbf q-\mbf Q)}
\otimes_{\mbf q}
P(\mbf q)
\overline{P(\mbf q+\mbf Q)}
\\
&=
C(\mbf q,\mbf Q)
\otimes_{\mbf q}
K(\mbf q,\mbf Q).
\end{align*}

For an ideal known probe, a possible reconstruction is obtained by applying an inverse Fourier transform over $\mbf q$ and performing division in the Wigner domain:
\begin{equation*}
C(\mbf q,\mbf Q)
=
\mathcal{F}_{\mbf r\to\mbf q}
\left(
\frac{
\mathcal{F}_{\mbf q\to\mbf r}^{-1}
\left(
I(\mbf q,\mbf Q)
\right)
}{
\mathcal{F}_{\mbf q\to\mbf r}^{-1}
\left(
K(\mbf q,\mbf Q)
\right)
}
\right).
\end{equation*}
In practice, the small correction $\epsilon>0$ is added to avoid instability in the division. Finally, $O(\mbf R)$ is estimated by evaluating the object cross-correlation at $\mbf q=\mbf 0$ and applying the conjugate form. Therefore, Wigner distribution deconvolution (WDD) provides a direct ptychographic reconstruction of $\overline{O(\mathbf R)}$ \cite{rodenburg1992theory}:
\begin{equation}
\mathcal{F}^{-1}_{\mathbf Q \to \mathbf R}
\left\{
\mathcal{F}_{\mathbf r\to\mathbf q}
\left[
\frac{
\mathcal{F}^{-1}_{\mathbf q\to\mathbf r}
\left\{
I(\mathbf q,\mathbf Q)
\right\}
\overline{
\mathcal{F}^{-1}_{\mathbf q\to\mathbf r}
\left\{
K(\mbf q,\mbf Q)
\right\}
}
}{
\card{
\mathcal{F}^{-1}_{\mathbf q\to\mathbf r}
\left\{
K(\mbf q,\mbf Q)
\right\}
}^2
+
\epsilon
}
\right]_{\mathbf q=\mathbf 0}
\right\}.
\label{eq:classic_deconv}
\end{equation}

Compared with model-based iterative approaches, WDD ptychography provides a fast direct reconstruction of the electrostatic potential over the region of interest (ROI) in 4D-STEM. Furthermore, prior work has shown that using a lower-dimensional Hermite--Gaussian representation can approximate the deconvolution efficiently, enabling real-time ptychographic reconstruction from 4D-STEM data
\cite{bangun2023wigner}.

Since HG functions are eigenfunctions of the Fourier transform, e.g.,
\begin{equation*}
\mcl{HG}_{\mbf r\to\mbf n}
\left\{
\mcl{F}^{-1}_{\mbf q\to\mbf r}
\left[
I(\mbf q,\mbf Q)
\right]
\right\}
=
i^{\card{\mbf n}}
\mcl{HG}_{\mbf q\to\mbf n}
\left\{
I(\mbf q,\mbf Q)
\right\},
\end{equation*}
they provide a compact representation of the reciprocal-space variable $\mbf q$. Although convolution is not exactly diagonal in the HG basis, a diagonal HG approximation maps the reciprocal-space variable $\mbf q$ to HG coefficients $\mbf n$.

Let $\mathcal S_M$ denote the set of $M$ retained two-dimensional HG modes. Under this diagonal HG approximation, the object $\widehat{O}(\mbf R)$ can be approximated by coefficient-wise division in HG space as
\begin{equation}
\mathcal{F}^{-1}_{\mathbf Q \to \mathbf R}
\left\{
\sum_{\mbf n\in\mathcal S_M}
\left[
\frac{
\mathcal{HG}_{\mathbf q\to\mathbf n}
\left\{
I(\mathbf q,\mathbf Q)
\right\}
\overline{
\mathcal{HG}_{\mathbf q\to\mathbf n}
\left\{
K(\mbf q,\mbf Q)
\right\}
}
}{
\card{
\mathcal{HG}_{\mathbf q\to\mathbf n}
\left\{
K(\mbf q,\mbf Q)
\right\}
}^2
+
\epsilon
}
\right]
\right\}.
\label{eq:low_deconv}
\end{equation}

Hence, instead of processing the full reciprocal-space representation, we project the detector-space data onto a lower-dimensional set of HG coefficients, thereby reducing memory and computational requirements. The reduced reconstruction assumes that the probe kernel used in the Wiener inversion is accurately known. In practice, inaccuracies in the estimated aberration coefficients lead to a mismatch between the reconstruction kernel and the ground-truth probe kernel, thereby introducing an additional reconstruction error.
 
\section{Main Results}
We analyze the reconstruction error arising from both the low-dimensional HG representation and inaccuracies in the estimated aberration coefficients. We further characterize the dependence of these errors on the Wiener regularization parameter and the measurement dose. The reconstruction error between the object
reconstructed using the low-dimensional aberration-corrected (AC) WDD under HG basis,
$O^{\mrm{AC}}_H(\mbf R)$, and that reconstructed using the classical WDD with
the ground-truth probe and Fourier transform, $O^{\mrm{GT}}_F(\mbf R)$:
\begin{equation}
\norm{
O^{\mrm{AC}}_H(\mbf R)
-
O^{\mrm{GT}}_F(\mbf R)
}_{L^2(\Omega_{\mbf R})}^2
\leq
\left(
C_{\mrm{hgf}}(M,\epsilon)
+
C_{\mrm{ab}}(M,\epsilon)
\norm{
\mathbf{c}^{\mrm{AC}}
-
\mathbf{c}^{\mrm{GT}}
}_2^2
\right)
D_{\max}D_{\mrm{tot}}.
\label{eq:main_bound}
\end{equation}
The reconstruction error consists of two contributions: the basis-mismatch
error between the truncated Hermite--Gaussian (HG) and Fourier
representations, and the error arising from inaccurate estimation of the
aberration coefficients. Here, $C_{\mrm{hgf}}(M,\epsilon)$ and
$C_{\mrm{ab}}(M,\epsilon)$ denote the corresponding constant given by basis-mismatch and
aberration-sensitivity coefficients, respectively. The bound explicitly
characterizes the dependence on the number of retained HG modes $M$, the
Wiener regularization parameter $\epsilon$, the aberration-coefficient
mismatch, and the dose factor $D_{\max}D_{\mrm{tot}}$. 
The key strategy to derive the bound is to introduce the low-dimensional
HG reconstruction obtained using the ground-truth aberration coefficients,
denoted by $O_H^{\mrm{GT}}(\mbf R)$, and add and subtract this quantity:
\begin{align}
O_H^{\mrm{AC}}(\mbf R)-O_F^{\mrm{GT}}(\mbf R)
&=
\left(
O_H^{\mrm{AC}}(\mbf R)-O_H^{\mrm{GT}}(\mbf R)
\right)
+
\left(
O_H^{\mrm{GT}}(\mbf R)-O_F^{\mrm{GT}}(\mbf R)
\right).
\label{eq:error_decomposition}
\end{align}
The first term isolates the reconstruction error arising from inaccuracies
in the estimated aberration coefficients while keeping the same
low-dimensional HG representation. The second term isolates the
HG--Fourier basis-mismatch error when the ground-truth probe is used in
both reconstructions. Consequently,
\begin{align*}
\norm{
O_H^{\mrm{AC}}(\mbf R)-O_F^{\mrm{GT}}(\mbf R)
}_{L^2(\Omega_{\mbf R})}^2
\leq\;
&
\underbrace{
2\norm{
O_H^{\mrm{AC}}(\mbf R)-O_H^{\mrm{GT}}(\mbf R)
}_{L^2(\Omega_{\mbf R})}^2
}_{\mathcal E_{\mrm{ab}}}
\nonumber
+
\underbrace{
2\norm{
O_H^{\mrm{GT}}(\mbf R)-O_F^{\mrm{GT}}(\mbf R)
}_{L^2(\Omega_{\mbf R})}^2
}_{\mathcal E_{\mrm{hgf}}}.
\label{eq:error_decomposition_bound}
\end{align*}
The first term ${\mathcal E_{\mrm{ab}}}$ isolates the reconstruction error arising from inaccuracies
in the estimated aberration coefficients while keeping the same
low-dimensional HG representation. The second term ${\mathcal E_{\mrm{hgf}}}$ isolates the
HG--Fourier basis-mismatch error when the ground-truth probe is used in
both reconstructions. The upper bounds are derived by applying the Cauchy--Schwarz inequality to separate the error terms, followed by Hölder's inequality and Parseval's identity. For the aberration-coefficient error, the Lipschitz bound of the Wiener map, the inequality $|e^{ix}-e^{iy}|\leq |x-y|$, and Bessel's inequality are additionally used to bound the probe difference, whereas the basis-mismatch error is controlled using the worst-case basis discrepancies and the corresponding Wiener-map bound. Finally, using $D(R)^2\leq D_{\max}D(R)$ and integrating over all scan positions yields  ${\mathcal E_{\mrm{ab}}}$ and ${\mathcal E_{\mrm{hgf}}}$. Complete derivation is given in supplementary materials.

Numerical experiments characterize these two error contributions separately.
For the HG--Fourier basis mismatch, an HG bandwidth of $\mrm{BW}=16$,
corresponding to $M=16^2=256$ retained two-dimensional modes, together with
a Wiener regularization parameter of approximately $\epsilon=10^{-3}$,
provides a practical low-error operating point. In particular, the relative
basis-mismatch error is approximately $10^{-3}$ or lower in this regime.
The aberration-induced error is governed primarily by
$\norm{\mathbf{c}^{\mrm{AC}}-\mathbf{c}^{\mrm{GT}}}_2^2$, and the numerical
results show that accurate estimation of the aberration coefficients is
therefore important for controlling the remaining reconstruction error.

\section{Numerical Experiments}
\begin{figure}[!htb] \centering \includegraphics[width=0.8\linewidth]{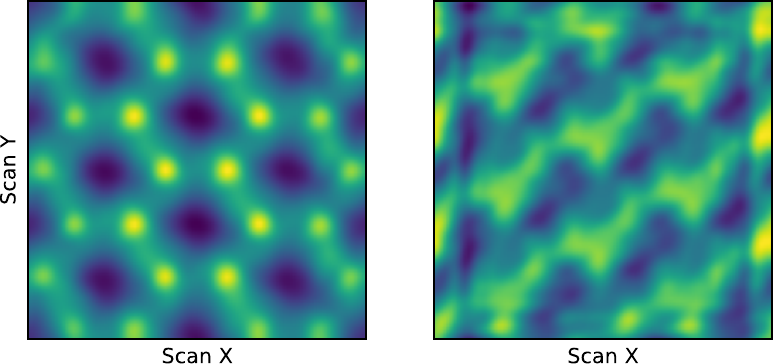} \caption{
WDD reconstruction of MoS$_2$ using an HG bandwidth of
$\mathrm{BW}=16$ ($M=256$) and $\epsilon=10^{-3}$.
Left: reconstruction using the ground-truth probe-aberration coefficients.
Right: reconstruction obtained by setting all aberration coefficients to zero.
}\label{fig:corr_vs_uncorr} \end{figure}

We conduct numerical experiments to validate the theoretical error decomposition and to characterize the sensitivity of the reconstruction to the main parameters appearing in the bound. In particular, we separately investigate the aberration-coefficient error and the HG--Fourier basis-mismatch error as functions of the aberration-coefficient mismatch, the Wiener regularization parameter, and the number of retained HG modes. In addition to these two deterministic error contributions, we also study the effect of electron dose by introducing Poisson noise into the simulated 4D-STEM measurements and evaluating the resulting reconstruction errors over multiple noise realizations.
\paragraph{Dataset Parameters : Electron Microscopy.} Table~\ref{tab:abtem_params} summarizes the abTEM \cite{madsen_abtem_2021} simulation parameters used to generate the MoS$_2$ dataset, including the simulation cell dimensions, spatial sampling, slice discretization, probe parameters, and aberration coefficients. Fine spatial sampling and infinite projection are used to efficiently model the periodic structure. The resulting 4D-STEM dataset has dimensions of $100 \times 100 \times 256 \times 256$, corresponding to a scan grid of $100 \times 100$ probe positions over a $10 \times 10~\text{\AA}^2$ scan area, with each diffraction pattern sampled on a $256 \times 256$ pixel grid. \begin{table}[!htb] \centering \caption{abTEM \cite{madsen_abtem_2021} simulation parameters and probe aberration coefficients used to generate the MoS$_2$ 4D-STEM dataset.} \begin{tabular}{lc@{\hspace{2em}}lc} \hline \multicolumn{2}{c}{\textbf{Simulation parameters}} & \multicolumn{2}{c}{\textbf{Aberration coefficients}} \\ \hline Parameter & Value & Coefficient & Value \\ \hline Cell size & $25.6 \times 25.6 \times 3.19~\text{\AA}^3$ & $C_{10}$ & $-79.14~\text{\AA}$ \\ Sampling & $0.1~\text{\AA}$ & $C_{12}$ & $22.65~\text{\AA}$ \\ Scan step & $0.1~\text{\AA}$ & $\phi_{12}$ & $-2.312~\mathrm{rad}$ \\ Number of slices & $2$ & $C_{21}$ & $257.60~\text{\AA}$ \\ Slice thickness & $1.595~\text{\AA}$ & $\phi_{21}$ & $0.842~\mathrm{rad}$ \\ Potential parametrization & Lobato \cite{lobato2014accurate} & $C_{23}$ & $643.07~\text{\AA}$ \\ Incident illumination & Focused electron probe & $\phi_{23}$ & $2.491~\mathrm{rad}$ \\ Electron energy & $200~\mathrm{keV}$ & $C_{30}$ & $-6.691\times10^{4}~\text{\AA}$ \\ \hline \multicolumn{2}{c}{} & \multicolumn{2}{l}{\footnotesize All other aberration coefficients are zero.} \end{tabular} \label{tab:abtem_params} \end{table} Figure~\ref{fig:corr_vs_uncorr} compares WDD reconstructions obtained with
and without aberration correction using an HG bandwidth of
$\mathrm{BW}=16$ ($M=256$) and $\epsilon=10^{-3}$.
When the reconstruction kernel incorporates the ground-truth aberration
coefficients, the MoS$_2$ structure is successfully recovered.
In contrast, setting all aberration coefficients to zero produces a
substantial degradation of the reconstruction and prevents accurate
recovery of the atomic structure. 

\paragraph{Dataset Parameters : X-Ray  Microscopy.}
To demonstrate the applicability of the proposed framework beyond electron
microscopy, we additionally generate synthetic coherent X-ray ptychography
data using the PyNX package~\cite{favre2020pynx}. The sample is represented by the built-in PyNX Siemens-star test object~\cite{favre2020pynx}. The simulated measurements consist of a regular $100\times100$ raster scan, with a $256\times256$ pixel far-field diffraction pattern recorded at each probe position, resulting in a four-dimensional dataset of size $100\times100\times256\times256$. The incident X-ray wavelength is set to $\lambda=1.5\times10^{-10}~\mathrm{m}$, the sample-to-detector distance to $1~\mathrm{m}$, and the detector pixel size to $55~\mu\mathrm{m}$, corresponding to an object-plane sampling of approximately $10.65~\mathrm{nm}$ per pixel. The illumination is generated using PyNX's focused-probe model with a circular aperture of radius $150~\mu\mathrm{m}$ and a focal length of $80~\mathrm{mm}$. A defocus of $350~\mu\mathrm{m}$ is introduced as the ground-truth probe aberration, while all other aberrations are set to zero. The raster step is approximately $31.96~\mathrm{nm}$, corresponding to a scan area of approximately $3.16\times3.16~\mu\mathrm{m}^{2}$. The initial experiments use noise-free diffraction intensities, while Poisson noise is introduced separately in the dose-dependent experiments.

\begin{figure}[!htb]
    \centering
    \includegraphics[width=0.9\linewidth]{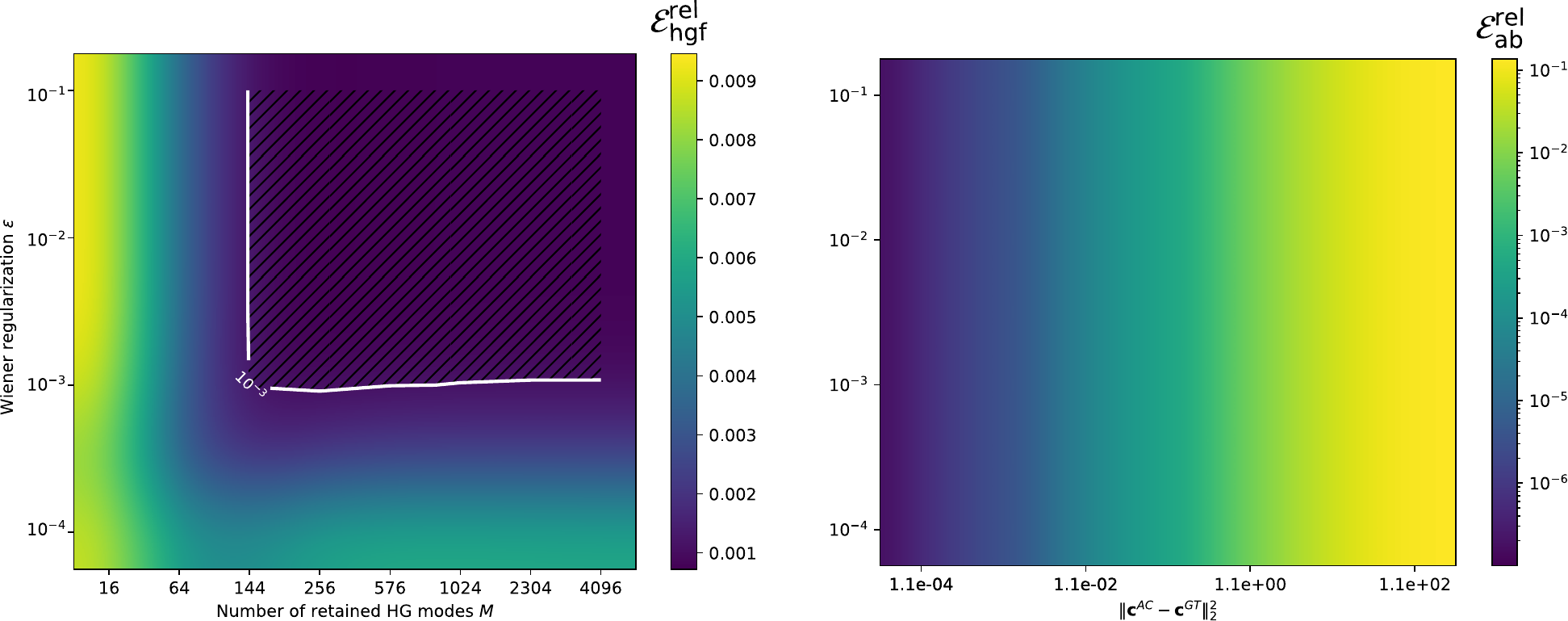}
    \caption{
    Numerical characterization of the two reconstruction-error contributions.
    (a) Relative HG--Fourier basis-mismatch error,
    $\mathcal{E}_{\rm hgf}^{\rm rel}$, as a function of the number of retained HG modes $M$ and the Wiener regularization parameter $\epsilon$. The hatched region denotes $\mathcal{E}_{\rm hgf}^{\rm rel}<10^{-3}$, with the white curve indicating the corresponding $10^{-3}$ contour.
    (b) Relative aberration-induced reconstruction error,
    $\mathcal{E}_{\rm ab}^{\rm rel}$, as a function of the squared aberration-coefficient mismatch $\norm{\mbf{c}^{\mrm{AC}}-\mbf{c}^{\mrm{GT}}}_2^2$ and $\epsilon$, evaluated at fixed HG bandwidth $\mathrm{BW}=16$ ($M=256$). The aberration-induced error is governed primarily by the coefficient mismatch and exhibits comparatively weak dependence on $\epsilon$ over the investigated range.
    }
    \label{fig:error_maps}
\end{figure}

\begin{figure}[!htb]
    \centering
    \includegraphics[width=0.8\linewidth]{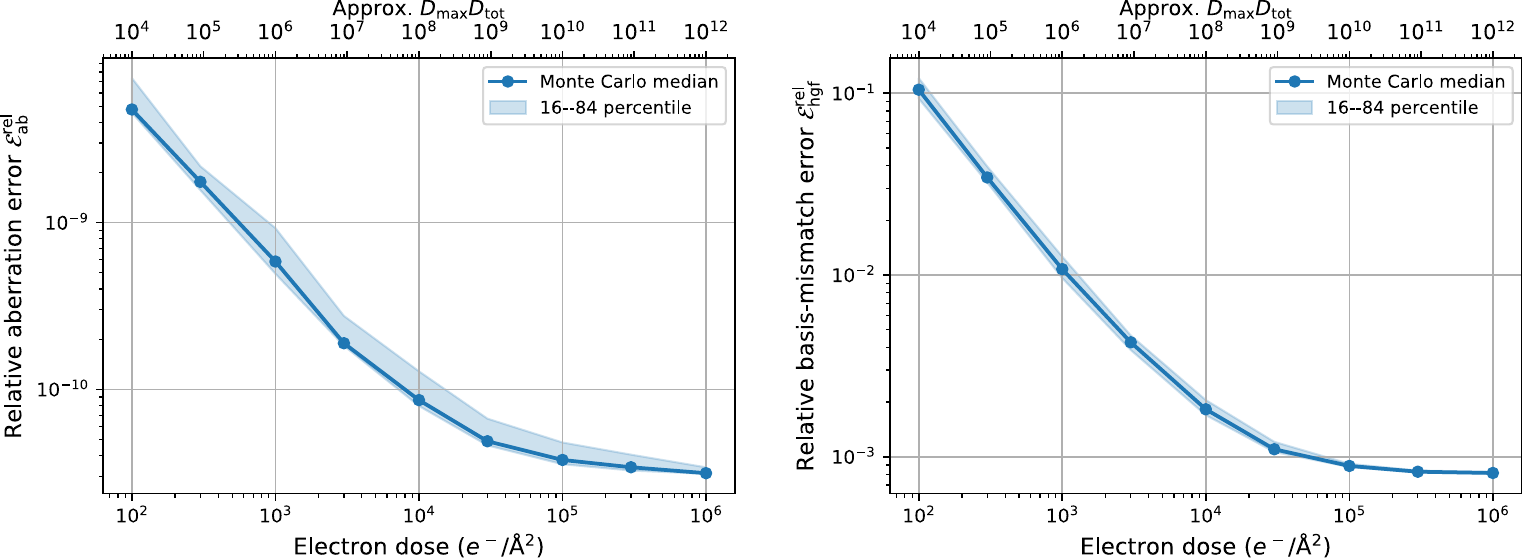}
    \caption{
    Dose dependence of the relative reconstruction errors under Poisson noise.
    (a) Relative aberration-induced reconstruction error,
    $\mathcal{E}_{\rm ab}^{\rm rel}$, evaluated using a fixed nonzero aberration-coefficient perturbation $\Delta\mbf{c}$.
    (b) Relative HG--Fourier basis-mismatch error,
    $\mathcal{E}_{\rm hgf}^{\rm rel}$, evaluated using the ground-truth aberration coefficients for both reconstructions.
    The horizontal axis denotes the mean electron dose in $e^-/\text{\AA}^2$, while the upper axis shows the approximate theoretical dose factor $D_{\max}D_{\rm tot}\approx D_{\text{\AA}^2}^2$ for the present $100\times100$ scan with $0.1\,\text{\AA}$ sampling.
    Solid curves indicate the median over Poisson-noise realizations, and the shaded regions denote the 16th--84th percentile range.
    Both error contributions decrease with increasing dose and approach high-dose residual floors, with the HG--Fourier basis mismatch providing the dominant residual contribution at high dose.
    }
    \label{fig:dose_errors}
\end{figure}

\begin{table}[!htb]
    \centering
    \caption{PyNX simulation parameters and probe aberration used to generate
    the synthetic X-ray ptychography dataset.}
    \begin{tabular}{lc@{\hspace{2em}}lc}
        \hline
        \multicolumn{2}{c}{\textbf{Simulation parameters}} &
        \multicolumn{2}{c}{\textbf{Probe aberration}} \\
        \hline
        Parameter & Value & Parameter & Value \\
        \hline

        Simulation package
        & PyNX~\cite{favre2020pynx}
        & Defocus
        & $350~\mu\mathrm{m}$ \\

        Object
        & Siemens star
        & Other aberrations
        & $0$ \\

        X-ray wavelength
        & $1.5\times10^{-10}~\mathrm{m}$
        & &
        \\

        Detector distance
        & $1~\mathrm{m}$
        & &
        \\

        Detector pixel size
        & $55~\mu\mathrm{m}$
        & &
        \\

        Object-plane sampling
        & $10.65~\mathrm{nm}$
        & &
        \\

        Incident illumination
        & Focused X-ray probe
        & &
        \\

        Aperture
        & Circular, $150~\mu\mathrm{m}$ radius
        & &
        \\

        Focal length
        & $80~\mathrm{mm}$
        & &
        \\

        Scan step
        & $31.96~\mathrm{nm}$
        & &
        \\

        \hline
    \end{tabular}
    \label{tab:pynx_params}
\end{table}
\begin{figure}[!htb] \centering \includegraphics[width=0.8\linewidth]{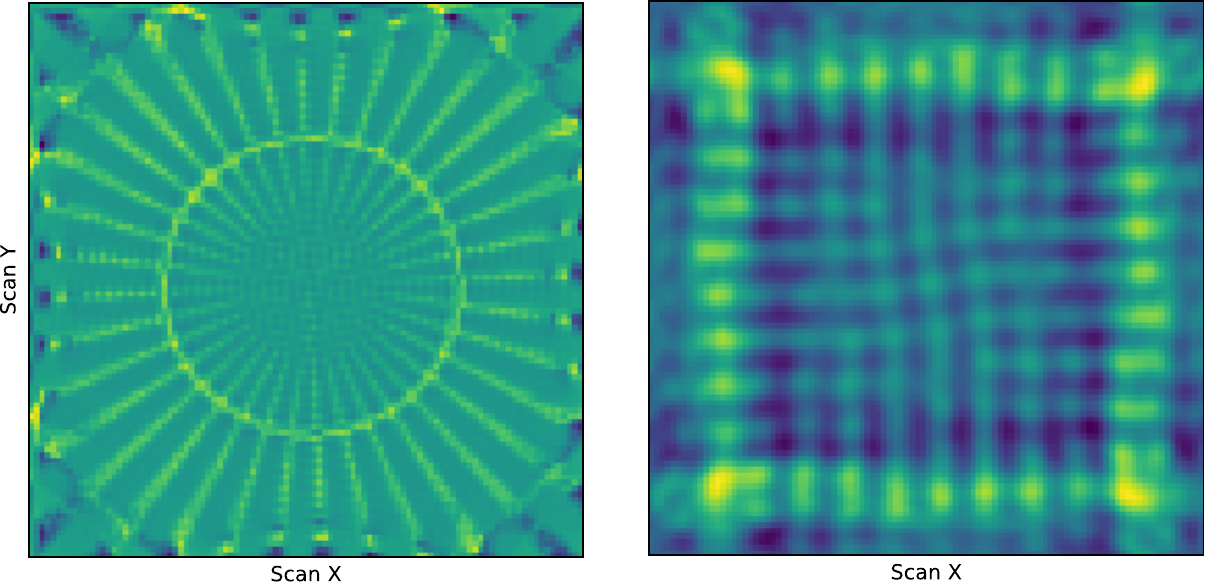} \caption{WDD reconstruction of Siemens Star object using an HG bandwidth of $16$ and $\epsilon=10^{-3}$. Left: reconstruction with the probe initialized using the ground-truth aberration coefficients. Right: reconstruction obtained by setting all aberration coefficients to zero.} \label{fig:corr_vs_uncorr_x} \end{figure}
\paragraph{Aberration-Coefficient and Basis-Mismatch Error.}
Numerical experiments were performed to separately characterize the aberration-induced and basis-mismatch contributions to the reconstruction error. For the aberration contribution, random perturbations of the aberration coefficients were applied along fixed Monte Carlo directions, and the resulting error was evaluated as a function of the coefficient mismatch $\norm{\mbf{c}^{\mrm{AC}}-\mbf{c}^{\mrm{GT}}}_2^2$ and the Wiener regularization parameter $\epsilon$, while the HG bandwidth was fixed at $16$ ($M=256$). For comparison across reconstruction settings, we report the corresponding
relative squared reconstruction errors, $
\mathcal E_{\rm ab}^{\rm rel}$ and $
\mathcal E_{\rm hgf}^{\rm rel}$. As shown in Fig.~\ref{fig:error_maps}(b), the aberration-induced reconstruction error increases strongly with the aberration-coefficient mismatch, whereas its dependence on $\epsilon$ is comparatively weak over the investigated range.
For the basis-mismatch contribution, the ground-truth aberration coefficients were used for both the HG and full Fourier reconstructions, thereby isolating the discrepancy introduced by the reduced HG representation. Figure~\ref{fig:error_maps}(a) shows the relative basis-mismatch error as a function of the number of retained HG modes $M$ and $\epsilon$. The error decreases rapidly as $M$ increases and reaches a low-error regime for sufficiently large $M$ and regularization. The contour $\mathcal{E}_{\rm hgf}^{\rm rel}=10^{-3}$ highlights the combinations of $M$ and $\epsilon$ for which the squared relative basis-mismatch error remains below this threshold. These results indicate that $M=256$ ($\mathrm{BW}=16$) provides a favorable accuracy--complexity trade-off and is therefore adopted in the subsequent experiments.
\paragraph{Dose Dependence of the Reconstruction Errors.}
To investigate the influence of measurement noise, we evaluated the reconstruction errors over a range of electron doses using Poisson-corrupted 4D-STEM intensities. For the present $100\times100$ scan with a real-space sampling of $0.1\,\text{\AA}$, the electron dose per scan position, $D$, and the areal dose, $D_{\text{\AA}^2}$, are related by $
D_{\text{\AA}^2}
=
\frac{D}{(0.1\,\text{\AA})^2}
=
100D. $
Under the approximately uniform-dose assumption,
$ 
D_{\rm tot}\approx 10^4D,
\,
D_{\max}\approx D,
$
and therefore
$ 
D_{\max}D_{\rm tot}
\approx
10^4D^2
=
D_{\text{\AA}^2}^2.
$
Thus, for the present scan geometry, the theoretical dose factor
$D_{\max}D_{\rm tot}$ can be related approximately to the square of the
areal electron dose. This correspondence is shown on the secondary horizontal
axis in Fig.~\ref{fig:dose_errors}. The secondary axis is included to relate
the experimental dose scale to the dose quantity appearing in the analytical
bound. Since the reconstructions reported below are evaluated using relative
errors under Poisson noise, the dose sweep is interpreted as a robustness
study rather than as a direct test of proportionality to
$D_{\max}D_{\rm tot}$.
Poisson noise was applied to the noiseless 4D-STEM intensities at mean electron doses ranging from $10^2$ to $10^6\,e^-/\text{\AA}^2$. For each dose level, multiple Poisson-noise realizations were generated, and the same realization was used for the HG and Fourier reconstructions being compared. 
The HG bandwidth, Wiener regularization parameter, probe, specimen, and scan geometry were kept fixed throughout the dose sweep. For the aberration contribution, a fixed nonzero coefficient perturbation, $
\Delta\mbf{c}
=
\mbf{c}^{\mrm{AC}}-\mbf{c}^{\mrm{GT}},$
was used at all dose levels. For the basis-mismatch contribution, both the HG and Fourier reconstructions used the ground-truth aberration coefficients, thereby isolating the HG--Fourier representation mismatch.

As shown in Fig.~\ref{fig:dose_errors}, both relative reconstruction errors
decrease with increasing electron dose and approach high-dose residual
floors. The aberration-induced contribution reaches a substantially lower
floor, whereas the HG--Fourier basis-mismatch error approaches a residual
level of approximately $10^{-3}$. This indicates that, for the investigated
reconstruction parameters, the basis mismatch dominates the residual
discrepancy in the high-dose regime.

\section{Conclusions}
We present a low-dimensional HG representation for direct WDD ptychographic reconstruction with explicit probe-aberration modeling. The reconstruction error relative to classical Fourier-domain WDD is decomposed into HG--Fourier basis-mismatch and aberration-coefficient  error, yielding an upper bound that depends on the retained HG modes, Wiener regularization, aberration mismatch, and measurement dose. Numerical experiments support this decomposition and characterize the corresponding parameter dependencies. The basis-mismatch error decreases rapidly with increasing HG dimension and reaches a low-error regime for suitable regularization, while the aberration-induced error is governed primarily by the accuracy of the aberration coefficients. Poisson-noise experiments further show that the relative errors decrease with increasing dose and approach high-dose residual floors. Simulated electron and X-ray reconstructions demonstrate that the framework is applicable to both electron- and photon-counting ptychographic settings. These results provide a theoretical and numerical basis for selecting compact HG representations and regularization parameters in efficient ptychographic reconstruction. Future work may extend the development of aberration-corrected algorithm for low-dimensional WDD.

\bibliographystyle{IEEEbib}
\bibliography{references}
\appendix
\section{Supplementary Materials}

\paragraph{Deriving the error bound.} We begin by comparing the object obtained from the aberration-corrected WDD reconstruction with the ground-truth WDD reconstruction.
\begin{equation}
\label{eq:originalE}
 {
\begin{aligned}
\norm{O^{\mathrm{AC}}_H(\mbf R) - {O}^{\mathrm{GT}}_F(\mbf R)}^2_{L^2(\Omega_{\mbf R})} &= \norm{\mcl{F}^{-1}_{\mbf Q \to \mbf R}\left(O^{\mathrm{AC}}_H(\mbf Q) \right) - \mcl{F}^{-1}_{\mbf Q \to \mbf R}\left({O}^{\mathrm{GT}}_F(\mbf Q)\right)} ^2_{L^2(\Omega_{\mbf R})}\\
&=\norm{O^{\mathrm{AC}}_H(\mbf Q)- {O}_F^{\mathrm{GT}}(\mbf Q)} ^2_{L^2(\Omega_{\mbf Q})}
\end{aligned}
}
\end{equation}
The last equation is derived from the Plancherel or Parseval property of Fourier transform. Thereby, we can directly write in terms of Fourier space of scan coordinates $\mbf Q$ both from \eqref{eq:classic_deconv} and \eqref{eq:low_deconv}.
\begin{equation}
    \begin{aligned}
        \int_{\Omega_Q}
\Bigg|
\sum_{\mathbf n\in\mathcal{S}_M}
\frac{
\mathcal{HG}_{\mathbf q\to\mathbf n}\{I(\mathbf q,\mathbf Q)\}
\overline{\mathcal{HG}_{\mathbf q\to\mathbf n}\{K^{\mathrm{AC}}(\mathbf q,\mathbf Q)\}}
}{
\big|\mathcal{HG}_{\mathbf q\to\mathbf n}\{K^{\mathrm{AC}}(\mathbf q,\mathbf Q)\}\big|^2+\epsilon
}
-
\int_{\Omega_r}
\frac{
\mathcal F^{-1}_{\mathbf q\to\mathbf r}\{I(\mathbf q,\mathbf Q)\}
\overline{\mathcal F^{-1}_{\mathbf q\to\mathbf r}\{K^{\mathrm{GT}}(\mathbf q,\mathbf Q)\}}
}{
\big|\mathcal F^{-1}_{\mathbf q\to\mathbf r}\{K^{\mathrm{GT}}(\mathbf q,\mathbf Q)\}\big|^2+\epsilon
}
\,d^2r
\Bigg|^2d^2Q,
    \end{aligned}
\end{equation}
where $I(\mathbf q,\mathbf Q)$ is the Fourier transform of the measured 4D intensity $I(\mbf q, \mbf R)$ over the scan coordinate $R$, and 
$K^{\mathrm{AC}}(\mathbf q,\mathbf Q)$ as well as 
$K^{\mathrm{GT}}(\mathbf q,\mathbf Q)$ denote the aberration-corrected and ground-truth kernels, respectively.
 
The integral over $\Omega_r$ appears as we want to perform Fourier transform from $\mbf r$ to $\mbf q$ and evaluate the result for $\mbf q = 0$. Since the HG sum is independent of $\mathbf r$ we can write
 
\begin{equation}
\norm{O_H^{\mathrm{AC}}(\mbf Q)- {O}_F^{\mathrm{GT}}(\mbf Q)} ^2_{L^2(\Omega_{\mbf Q})}
=
\int_{\Omega_Q}
\left|
\int_{\Omega_r}F(\mathbf r,\mathbf Q)\,d^2r
\right|^2d^2Q,
\end{equation}
where $
F(\mathbf r,\mathbf Q) $ is given as
\begin{equation}
\label{eq:Fdef}
\begin{aligned}
&\frac{1}{|\Omega_r|}
\sum_{\mathbf n\in\mathcal S_M}
\frac{
\mathcal{HG}_{\mathbf q\to\mathbf n}\{I(\mathbf q,\mathbf Q)\}
\overline{\mathcal{HG}_{\mathbf q\to\mathbf n}\{K^{\mathrm{AC}}(\mathbf q,\mathbf Q)\}}
}{
\big|\mathcal{HG}_{\mathbf q\to\mathbf n}\{K^{\mathrm{AC}}(\mathbf q,\mathbf Q)\}\big|^2+\epsilon
} -
\frac{
\mathcal F^{-1}_{\mathbf q\to\mathbf r}\{I(\mathbf q,\mathbf Q)\}
\overline{\mathcal F^{-1}_{\mathbf q\to\mathbf r}\{K^{\mathrm{GT}}(\mathbf q,\mathbf Q)\}}
}{
\big|\mathcal F^{-1}_{\mathbf q\to\mathbf r}\{K^{\mathrm{GT}}(\mathbf q,\mathbf Q)\}\big|^2+\epsilon
}.
\end{aligned}
\end{equation}
Inserting the HG for the ground truth we can decompose the error in terms of aberration and differences between Fourier and HG basis, namely $F(\mathbf r,\mathbf Q) = F_{\mathrm{ab}}(\mathbf r,\mathbf Q) + F_{\mathrm{hgf}}(\mathbf r,\mathbf Q)$, where using Cauchy--Schwarz in $\mathbf r$ we have,
\begin{equation}
\begin{aligned}
\left|\int_{\Omega_r}F(\mathbf r,\mathbf Q)\,d^2r\right|^2
&\le
|\Omega_r|\int_{\Omega_r}|F(\mathbf r,\mathbf Q)|^2\,d^2r = |\Omega_r|\int_{\Omega_r}|F_{\mathrm{ab}}(\mathbf r,\mathbf Q) + F_{\mathrm{hgf}}(\mathbf r,\mathbf Q)|^2\,d^2r \\
&\le 2|\Omega_r|\int_{\Omega_r}|F_{\mathrm{ab}}(\mathbf r,\mathbf Q)|^2\,d^2r + 2|\Omega_r|\int_{\Omega_r}| F_{\mathrm{hgf}}(\mathbf r,\mathbf Q)|^2\,d^2r,
\end{aligned}
\end{equation}
where $F_{\mathrm{ab}}(\mathbf r,\mathbf Q)$ is given as
\begin{equation}
\label{eq:Fab}
\frac{1}{|\Omega_r|}
\sum_{\mathbf n\in\mathcal S_M}
\mathcal{HG}_{\mathbf q\to\mathbf n}\{I(\mathbf q,\mathbf Q)\}
\left[
W_\epsilon\!\left(\mathcal{HG}_{\mathbf q\to\mathbf n}\{K^{\mathrm{AC}}(\mathbf q,\mathbf Q)\}\right)
-
W_\epsilon\!\left(\mathcal{HG}_{\mathbf q\to\mathbf n}\{K^{\mathrm{GT}}(\mathbf q,\mathbf Q)\}\right)
\right],
\end{equation}
and $F_{\mathrm{hgf}}(\mathbf r,\mathbf Q)$ is written as:
\begin{equation}
\label{eq:Fhgf}
\begin{aligned}
&\frac{1}{|\Omega_r|}
\sum_{\mathbf n\in\mathcal S_M}
{
\mathcal{HG}_{\mathbf q\to\mathbf n}\{I(\mathbf q,\mathbf Q)\}
W_\epsilon\!\left( {\mathcal{HG}_{\mathbf q\to\mathbf n}\{K^{\mathrm{GT}}(\mathbf q,\mathbf Q)\}} \right)
}
-
{\mathcal F^{-1}_{\mathbf q\to\mathbf r}\{I(\mathbf q,\mathbf Q)\}
W_\epsilon\!\left({\mathcal F^{-1}_{\mathbf q\to\mathbf r}\{K^{\mathrm{GT}}(\mathbf q,\mathbf Q)\}}\right)
}.
\end{aligned}
\end{equation}

Here we follow the notation from  Wiener map $
W_\epsilon(z):=\frac{\overline z}{|z|^2+\epsilon},$
where $\epsilon>0$. Hence, the error can be decomposed into aberrated-correction error in HG representation and the mismatch between Fourier and HG representation. In the following, we focus on deriving each component of the error, namely aberration-coefficient error and the basis mismatch error.
\paragraph{Aberration-coefficient error.}

\begin{figure}[!htb]
\centering
\includegraphics[width=0.82\textwidth]{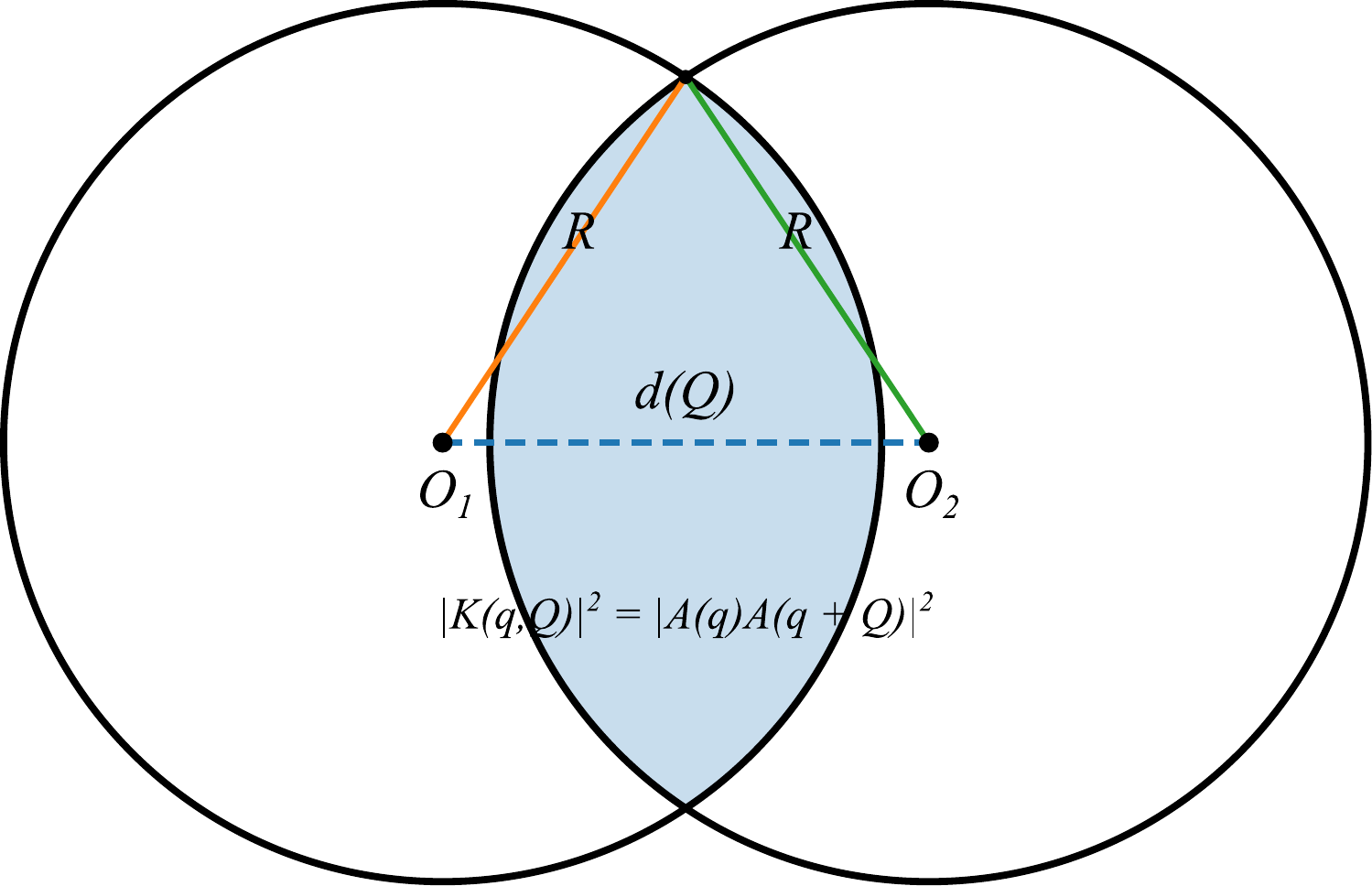}
\caption{Geometry of the trotter/overlap function for a hard circular pupil $A(\mbf q)$. Two disks of radius $R$ are separated by shift distance $d(\mbf Q)$.}
\label{fig:trotter}
\end{figure}

Using the Lipschitz bound of $W_\epsilon$ in  and Cauchy--Schwarz in the finite index set we can upper bound the $|F_{\mathrm{ab}}(\mathbf r,\mathbf Q)|^2$ as,
\begin{equation}
\label{eq:EabFirst}
\begin{aligned}
&
\frac{1}{|\Omega_r|^2\epsilon^2}
\left(
\sum_{\mathbf n\in\mathcal S_M}
\big|\mathcal{HG}_{\mathbf q\to\mathbf n}\{I(\mathbf q,\mathbf Q)\}\big|^2
\right)
\left(
\sum_{\mathbf n\in\mathcal S_M}
\big|\mathcal{HG}_{\mathbf q\to\mathbf n}\{K^{\mathrm{AC}}(\mathbf q,\mathbf Q)\}
-
\mathcal{HG}_{\mathbf q\to\mathbf n}\{K^{\mathrm{GT}}(\mathbf q,\mathbf Q)\}\big|^2
\right).
\end{aligned}
\end{equation}
Write the probe formation in reciprocal space, we can have the correlation $K( \mbf q, \mbf Q)$ in reciprocal space \eqref{eq:probe_recip} as
\begin{equation}
K(\mathbf q,\mathbf Q)=\card{A(\mbf q)A(\mbf q + \mbf Q)}e^{i\left[\chi(\mbf q) + \chi(\mbf q + \mbf Q)\right]} ,
\end{equation}
and the aberration function can be expanded in aberration basis function
\begin{equation*}
\chi(\mbf q) + \chi(\mbf q + \mbf Q) = \frac{2\pi}{\lambda}
\sum_{l=1}^L c_l\left[\Psi_l(\mbf q) + \Psi_l(\mbf q + \mbf Q) \right]
\end{equation*}
Thereby, the difference between the estimation and ground truth can be written as 
\begin{equation}
\left(\chi(\mbf q) + \chi(\mbf q + \mbf Q)\right)^{\mathrm{AC}}-\left( \chi(\mbf q) + \chi(\mbf q + \mbf Q) \right)^{\mathrm{GT}}
=
\frac{2\pi}{\lambda}\sum_{l=1}^L\left(c_l^{\mrm{AC}} - c_l^{\mrm{GT}} \right)\left[\Psi_l(\mbf q) + \Psi_l(\mbf q + \mbf Q) \right].
\end{equation}
Using $|e^{ix}-e^{iy}|\le |x-y|$ and the Bessel innequality of orthogonal HG projection \cite{kreyszig1991introductory}. Since the Hermite--Gaussian modes form an orthonormal system in
$L^2(\Omega_q)$, Bessel's inequality yields
\[
\sum_{\mathbf n\in S_M}
\left|
\mathcal H_{\mathbf q\to\mathbf n}\{f(\mathbf q)\}
\right|^2
\leq
\|f\|_{L^2(\Omega_q)}^2
=
\int_{\Omega_q}|f(\mathbf q)|^2\,d^2q. 
\] Hence, we can derive the upper bound as follows:
\begin{equation}
\label{eq:KdiffBound}
\sum_{\mathbf n\in\mathcal S_M}
\big|\mathcal{HG}_{\mathbf q\to\mathbf n}\{K^{\mathrm{AC}}(\mathbf q,\mathbf Q)
-
 K^{\mathrm{GT}}(\mathbf q,\mathbf Q)\}\big|^2
\le
\int_{\Omega_q}|A(\mbf q)A(\mbf q + \mbf Q)|^2
\left|\sum_{l=1}^L\Delta c_l Z_l(\mathbf q, \mbf Q)\right|^2d^2q,
\end{equation}
where $Z_l(\mathbf q, \mbf Q)=\Psi_l(\mbf q) + \Psi_l(\mbf q + \mbf Q)$. Applying Cauchy--Schwarz to the finite sum over aberration
coefficients gives
\begin{align}
\left|
\sum_{l=1}^{L}
\Delta c_l
Z_l(\mathbf q,\mathbf Q)
\right|^2
&\le
\left(
\sum_{l=1}^{L}|\Delta c_l|^2
\right)
\left(
\sum_{l=1}^{L}
|Z_l(\mathbf q,\mathbf Q)|^2
\right)
\nonumber =
\|\Delta\mathbf c\|_2^2
\sum_{l=1}^{L}
|Z_l(\mathbf q,\mathbf Q)|^2 .
\label{eq:cauchy_aberration}
\end{align}
Therefore we can write the difference as
\begin{align}
\sum_{\mathbf n\in\mathcal S_M}
\big|\mathcal{HG}_{\mathbf q\to\mathbf n}\{K^{\mathrm{AC}}(\mathbf q,\mathbf Q)
-
 K^{\mathrm{GT}}(\mathbf q,\mathbf Q)\}\big|^2
&\le
\left(\frac{2\pi}{\lambda}\right)^2
\|\Delta\mathbf c\|_2^2
\int_{\Omega_q}
|A(\mathbf q)A(\mathbf q+\mathbf Q)|^2
\sum_{l=1}^{L}
|Z_l(\mathbf q,\mathbf Q)|^2
\,d^2q .
\end{align}
Using H\"older's inequality,
\begin{align}
&\int_{\Omega_q}
|A(\mathbf q)A(\mathbf q+\mathbf Q)|^2
\sum_{l=1}^{L}
|Z_l(\mathbf q,\mathbf Q)|^2
\,d^2q
\nonumber
&\le
\left[
\max_{\mathbf q\in\Omega_{\mathrm{ov}}(\mathbf Q)}
\sum_{l=1}^{L}
|Z_l(\mathbf q,\mathbf Q)|^2
\right]
\int_{\Omega_q}
|A(\mathbf q)A(\mathbf q+\mathbf Q)|^2
\,d^2q .
\label{eq:holder_aberration}
\end{align}
For commonly used binary circular pupil $  
A(\mathbf q)
=
\mathbf 1_{\{|\mathbf q|\le R\}}$,
and hence $
|A(\mathbf q)A(\mathbf q+\mathbf Q)|^2
=
\mathbf 1_{\Omega_{\mathrm{ov}}(\mathbf Q)}(\mathbf q).$
Consequently$
\int_{\Omega_q}
|A(\mathbf q)A(\mathbf q+\mathbf Q)|^2
\,d^2q
=
|\Omega_{\mathrm{ov}}(\mathbf Q)|
\label{eq:overlap_measure}
$. For two circular pupils of radius $R$, $
|\Omega_{\mathrm{ov}}(\mathbf Q)|
\le
\pi R^2$
with equality at $\mathbf Q=0$ shown in Figure \ref{fig:trotter}.
Thus,
\begin{equation}
\sum_{\mathbf n\in\mathcal S_M}
\big|\mathcal{HG}_{\mathbf q\to\mathbf n}\{K^{\mathrm{AC}}(\mathbf q,\mathbf Q)\}
-
 K^{\mathrm{GT}}(\mathbf q,\mathbf Q)\}\big|^2 \le
\pi R^2 \left(\frac{2\pi}{\lambda}\right)^2
\|\Delta\mathbf c\|_2^2
\max_{\mathbf q\in\Omega_{\mathrm{ov}}(\mathbf Q)}
\sum_{l=1}^{L}
|Z_l(\mathbf q,\mathbf Q)|^2.
\label{eq:intersect}
\end{equation}
In the continuous reciprocal-space formulation,
$R =
\frac{\sin\alpha}{\lambda}
\simeq
\frac{\alpha}{\lambda},
 $
where $\alpha$ is the probe semi-convergence angle and $\lambda$ is the wavelength. Therefore,
 $
\pi R^2
=
\frac{\pi\sin^2\alpha}{\lambda^2}
\simeq
\frac{\pi\alpha^2}{\lambda^2}.
$
For the Krivanek aberration basis \eqref{eq:krivanek}, the individual basis functions
may be written as
\begin{equation}
\Psi_{nm,a}(q,\phi)
=
\frac{(\lambda q)^{n+1}}{n+1}
\cos(m\phi) \quad
\text{and}
\quad
\Psi_{nm,b}(q,\phi)
=
\frac{(\lambda q)^{n+1}}{n+1}
\sin(m\phi).
\end{equation}
Inside the pupil $\lambda q\le\alpha$
and therefore $
|\Psi_l(\mathbf q)|
\le
\frac{\alpha^{n_l+1}}{n_l+1} $. Since both $\mathbf q$ and $\mathbf q+\mathbf Q$ lie inside the
overlap region,
\begin{align}
|Z_l(\mathbf q,\mathbf Q)|
&=
|\Psi_l(\mathbf q)+\Psi_l(\mathbf q+\mathbf Q)|
\nonumber\le
|\Psi_l(\mathbf q)|
+
|\Psi_l(\mathbf q+\mathbf Q)|
\nonumber \le
\frac{2\alpha^{n_l+1}}{n_l+1}.
\end{align}
Hence, summing over the $L$ corrected aberration coefficients gives
\begin{equation}
\max_{\mathbf q\in\Omega_{\mathrm{ov}}(\mathbf Q)}
\sum_{l=1}^{L}
|Z_l(\mathbf q,\mathbf Q)|^2
\le
4
\sum_{l=1}^{L}
\frac{\alpha^{2(n_l+1)}}{(n_l+1)^2}.
\label{eq:kriv_sum_bound}
\end{equation}
Combining  
\eqref{eq:intersect} and \eqref{eq:kriv_sum_bound} yields
\begin{align}
&\sum_{\mathbf n\in\mathcal S_M}
\big|\mathcal{HG}_{\mathbf q\to\mathbf n}\{K^{\mathrm{AC}}(\mathbf q,\mathbf Q) 
-
 K^{\mathrm{GT}}(\mathbf q,\mathbf Q)\}\big|^2 
\nonumber\le
\frac{\alpha^2 16\pi^3}{\lambda^4}
\|\Delta\mathbf c\|_2^2
\sum_{l=1}^{L}
\frac{\alpha^{2n_l+2}}{(n_l+1)^2}.
\end{align}
It should be noted that,we assume \(0\le\alpha\le1\) rad, which covers the physically relevant semi-convergence angles considered here.   Thereby, the finite sum can be written explicitly in terms of the aberration order \(n\) as
$
S_N(\alpha)
=
\sum_{n=0}^{N}
g_n\frac{\alpha^{2(n+1)}}{(n+1)^2},
$
where \(g_n\) denotes the number of aberration coefficients of order \(n\).
Setting
$
k=n+1,
$
gives $
S_N(\alpha)
=
\sum_{k=1}^{N+1}
g_{k-1}\frac{\alpha^{2k}}{k^2}.$
Let $
g_{\max}=\underset{0\leq n\leq N}{\max} g_n$ is the maximum number of aberration coefficients of all orders. For \(0\leq\alpha\leq1\), we have $
\alpha^{2k}\leq\alpha^2 $
and therefore $
S_N(\alpha)
\leq
g_{\max}\alpha^2
\sum_{k=1}^{N+1}\frac{1}{k^2}
=
g_{\max}\alpha^2 H_{N+1}^{(2)}
\leq
g_{\max}\frac{\pi^2}{6}\alpha^2,$
where \(H_N^{(2)}\) denotes the generalized harmonic number of order \(2\). From the original inequality in \eqref{eq:EabFirst} we write:
\begin{align}
|F_{\mathrm{ab}}(\mathbf r, \mathbf Q)|^2
\le
\frac{1}{\card{\Omega_{\mbf r}}^2\epsilon^2}
&
\left[
\sum_{\mathbf n\in\mathcal S_M}
\left|
\mathcal{HG}_{\mathbf q\to\mathbf n}\{I(\mathbf q,\mathbf Q)\}
\right|^2
\right]
\nonumber \times
\frac{\alpha^2 16\pi^3}{\lambda^4}
\|\Delta\mathbf c\|_2^2´\,
g_{\max} \frac{\pi^2}{6} \alpha^2,
\label{eq:F_ab_cs}
\end{align}
Define the dose at scan position $\mathbf R$ by
$ 
D(\mathbf R)
:=
\int_{\Omega_q}
I(\mathbf q,\mathbf R),d^2q.
$
Then, by H\"older's inequality,
\begin{equation}
\left|
\mathcal{HG}_{\mathbf q\to\mathbf n}
\{I(\mathbf q,\mathbf R)\}
\right|
\le
D(\mathbf R)
\|\Phi_{\mathbf n}\|_{L^\infty(\Omega_q)}.
\end{equation}
Hence,
\begin{equation}
\sum_{\mathbf n\in\mathcal S_M}
\left|
\mathcal{HG}_{\mathbf q\to\mathbf n}
\{I(\mathbf q,\mathbf R)\}
\right|^2
\le
D(\mathbf R)^2
\sum_{\mathbf n\in\mathcal S_M}
\|\Phi_{\mathbf n}\|_{L^\infty(\Omega_q)}^2.
\end{equation}
For the standard separable two-dimensional $L^2$-normalized HG basis,
$
\|\Phi_{\mathbf n}\|_{L^\infty(\Omega_q)}
\le \pi^{-1/2},
$
and therefore
$
\sum_{\mathbf n\in\mathcal S_M}
\|\Phi_{\mathbf n}\|_{L^\infty(\Omega_q)}^2
\le \frac{M}{\pi},
$ where $M$ is the cardinality of set $\mcl{S}_M$. Consequently,
\begin{align}
\mathcal E_{\mathrm{ab}}
&\le
\frac{1}{\epsilon^2}
\left[
\int_{\Omega_Q}
\sum_{\mathbf n\in\mathcal S_M}
\left|
\mathcal{HG}_{\mathbf q\to\mathbf n}
\{I(\mathbf q,\mathbf Q)\}
\right|^2
\,d^2Q
\right]
\frac{32\pi^5}{6\lambda^4}
g_{\max}\alpha^4
\|\Delta\mathbf c\|_2^2
\\
&=
\frac{1}{\epsilon^2}
\left[
\int_{\Omega_R}
\sum_{\mathbf n\in\mathcal S_M}
\left|
\mathcal{HG}_{\mathbf q\to\mathbf n}
\{I(\mathbf q,\mathbf R)\}
\right|^2
\,d^2R
\right]
\frac{32\pi^5}{6\lambda^4}
g_{\max}\alpha^4
\|\Delta\mathbf c\|_2^2
\\
&\le
\frac{1}{\epsilon^2}
\left[
\frac{M}{\pi}
\int_{\Omega_R}
D(\mathbf R)^2\,d^2R
\right]
\frac{32\pi^5}{6\lambda^4}
g_{\max}\alpha^4
\|\Delta\mathbf c\|_2^2
\\
&\le
C_{\mathrm{ab}}(M,\epsilon)
D_{\max}D_{\mathrm{total}}
\|\Delta\mathbf c\|_2^2.
\end{align}
where $D_{\mathrm{max}}=\underset{\mathbf R}{\max} \,D(\mathbf R)$ denotes the
maximum dose over all scan positions and
$D_{\mathrm{total}}=\int_{\Omega_R}D(\mathbf R)\,d^2R$ denotes the
total dose. The second equality follows by commuting the HG transform with the Fourier transform in the scan coordinate and applying Parseval's identity in the conjugate variables \(\mathbf R\leftrightarrow\mathbf Q\). For fixed experimental settings, $g_{\max}, \alpha$ and $\lambda$
are constants and are therefore absorbed into
$C_{\mathrm{ab}}(M,\epsilon)$. The dependence on $M$ and $\epsilon$
is investigated in the numerical experiments.
 
\paragraph{Basis mismatch error.} Apart from the aberration error, here we also define the basis mismatch error since we perform HG representation. Let $
c_r:=\frac{|\Omega_r|}{M}$ and $ 
M=\operatorname{Card}(\mathcal S_M),
$  the basis mismatch error $F_{\mathrm{hgf}}(\mathbf r,\mathbf Q)$: \begin{align}
F_{\mathrm{hgf}}(\mathbf r,\mathbf Q) &= \frac{1}{|\Omega_r|}
\sum_{\mathbf n\in\mathcal S_M}
W_\epsilon\!\left(
\mathcal{HG}_{\mathbf q\to\mathbf n}\{K^{GT}(\mathbf q,\mathbf Q)\}
\right)
\left[
\mathcal{HG}_{\mathbf q\to\mathbf n}\{I(\mathbf q,\mathbf Q)\}
-
c_r\mathcal F^{-1}_{\mathbf q\to\mathbf r}\{I(\mathbf q,\mathbf Q)\}
\right]
\nonumber\\
&+\frac{1}{|\Omega_r|}
\sum_{\mathbf n\in\mathcal S_M}
c_r\mathcal F^{-1}_{\mathbf q\to\mathbf r}\{I(\mathbf q,\mathbf Q)\}
\left[
W_\epsilon\!\left(
\mathcal{HG}_{\mathbf q\to\mathbf n}\{K^{GT}(\mathbf q,\mathbf Q)\}
\right)
-
W_\epsilon\!\left(
\mathcal F^{-1}_{\mathbf q\to\mathbf r}\{K^{GT}(\mathbf q,\mathbf Q)\}
\right)
\right],
\end{align}
Using Cauchy-Schwarz inequality and $
|a+b|^2\le 2|a|^2+2|b|$ as well as upper bound of Wiener maps $W_\epsilon$: 
\begin{align*}
|F_{\mathrm{hgf}}(\mathbf r,\mathbf Q)|^2
&\le
\frac{M}{|\Omega_r|^2}
\frac{1}{2\epsilon}\sum_{\mathbf n\in\mathcal S_M}
\left|
\mathcal{HG}_{\mathbf q\to\mathbf n}\{I(\mathbf q,\mathbf Q)\}
-
c_r\mathcal F^{-1}_{\mathbf q\to\mathbf r}\{I(\mathbf q,\mathbf Q)\}
\right|^2
\nonumber\\
&+\frac{M}{|\Omega_r|^2}\frac{2c_r^2}{\epsilon^2}
\sum_{\mathbf n\in\mathcal S_M}
\left|
\mathcal F^{-1}_{\mathbf q\to\mathbf r}\{I(\mathbf q,\mathbf Q)\}
\right|^2
\left|
\mathcal{HG}_{\mathbf q\to\mathbf n}\{K^{GT}(\mathbf q,\mathbf Q)\}
-
\mathcal F^{-1}_{\mathbf q\to\mathbf r}\{K^{GT}(\mathbf q,\mathbf Q)\}
\right|^2 .
\end{align*}
Therefore, we write the total error $
\mathcal E_{\mathrm{hgf}}
:=
2|\Omega_r|
\int_{\Omega_r}\int_{\Omega_Q}
|F_{\mathrm{hgf}}(\mathbf r,\mathbf Q)|^2
\,d^2Q\,d^2r .
$
The first difference can be written as
\begin{align}
&\mathcal{HG}_{\mathbf q\to\mathbf n}
\{I(\mathbf q,\mathbf Q)\}
-
c_r\mathcal F^{-1}_{\mathbf q\to\mathbf r}
\{I(\mathbf q,\mathbf Q)\}
\nonumber =
\int_{\Omega_q}
I(\mathbf q,\mathbf Q)
\left[
\Phi_{\mathbf n}(\mathbf q)
-
c_r e^{\,i2\pi\mathbf q\cdot\mathbf r}
\right]
\,d^2q .
\end{align}
Since the HG transform and the inverse Fourier transform in $\mathbf q$
commute with the Fourier transform
$\mathbf R\leftrightarrow\mathbf Q$, Parseval gives
\begin{align}
&\int_{\Omega_Q}
\left|
\mathcal{HG}_{\mathbf q\to\mathbf n}
\{I(\mathbf q,\mathbf Q)\}
-
c_r\mathcal F^{-1}_{\mathbf q\to\mathbf r}
\{I(\mathbf q,\mathbf Q)\}
\right|^2
\,d^2Q
\nonumber =
\int_{\Omega_R}
\left|
\int_{\Omega_q}
I(\mathbf q,\mathbf R)
\left[
\Phi_{\mathbf n}(\mathbf q)
-
c_r e^{\,i2\pi\mathbf q\cdot\mathbf r}
\right]
\,d^2q
\right|^2
\,d^2R .
\end{align}
First, define $D(\mathbf R)
=
\int_{\Omega_q}
I(\mathbf q,\mathbf R)\,d^2q$ by H\"older's inequality
\begin{align}
\int_{\Omega_Q}
\left|
\mathcal{HG}_{\mathbf q\to\mathbf n}\{I(\mathbf q,\mathbf Q)\}
-
c_r\mathcal F^{-1}_{\mathbf q\to\mathbf r}\{I(\mathbf q,\mathbf Q)\}
\right|^2
\,d^2Q \le
\|\Phi_{\mathbf n}(\mathbf q)
-
c_r e^{\,i2\pi\mathbf q\cdot\mathbf r}\|_{L^\infty(\Omega_{\mbf q})}^2
\int_{\Omega_R}D(\mathbf R)^2\,d^2R
\label{eq:first}
\end{align}
Similarly, for the second equation, applying Hölder inequality and use the fact that maximum absolute value Fourier basis is unit modulus, we arrive at
\begin{align}
\int_{\Omega_Q}
\left|
\mathcal F^{-1}_{\mathbf q\to\mathbf r}\{I(\mbf q, \mbf Q)\}
\right|^2
\,d^2Q
&=
\int_{\Omega_R}
\left|
\int_{\Omega_q}
I(\mathbf q,\mathbf R)e^{\,i2\pi\mathbf q\cdot\mathbf r} d^2q
\right|^2
\,d^2R \le
\int_{\Omega_R}D(\mathbf R)^2\,d^2R ,
\label{eq:next}
\end{align} 
For the kernel difference, H\"older's inequality gives
\begin{align}
&
\left|
\mathcal{HG}_{\mathbf q\to\mathbf n}
\{K^{GT}(\mathbf q,\mathbf Q)\}
-
\mathcal F^{-1}_{\mathbf q\to\mathbf r}
\{K^{GT}(\mathbf q,\mathbf Q)\}
\right|  \le
\left(
\int_{\Omega_q}
\left|K^{GT}(\mathbf q,\mathbf Q)\right|
\,d^2q
\right)
\left\|
\Phi_{\mathbf n}(\mathbf q)
-
e^{\,i2\pi\mathbf q\cdot\mathbf r}
\right\|_{L^\infty(\Omega_q)}.
\label{eq:second}
\end{align}
The overlap of the probe is given by
$\int_{\Omega_q}
\left|K^{GT}(\mathbf q,\mathbf Q)\right|
\,d^2q
=
|\Omega_{\mathrm{ov}}(\mathbf Q)|
\le
\pi R^2
\simeq
\frac{\pi\alpha^2}{\lambda^2}.$
Combining  \eqref{eq:first}, \eqref{eq:next}, and \eqref{eq:second}  we now integrate the pointwise bound over
$\mathbf r\in\Omega_r$ and $\mathbf Q\in\Omega_Q$. Recalling the definition
of the total basis-mismatch contribution,
\begin{equation}
\mathcal E_{\mathrm{hgf}}
:=
2|\Omega_r|
\int_{\Omega_r}\int_{\Omega_Q}
|F_{\mathrm{hgf}}(\mathbf r,\mathbf Q)|^2
\,d^2Q\,d^2r ,
\end{equation}
we obtain
\begin{align}
\mathcal E_{\mathrm{hgf}}
&\le{}
\frac{M}{|\Omega_r|\epsilon}
\sum_{\mathbf n\in\mathcal S_M}
\int_{\Omega_r}\int_{\Omega_Q}
\left|
\mathcal{HG}_{\mathbf q\to\mathbf n}\{I(\mathbf q,\mathbf Q)\}
-
c_r\mathcal F^{-1}_{\mathbf q\to\mathbf r}
\{I(\mathbf q,\mathbf Q)\}
\right|^2
\,d^2Q\,d^2r
\nonumber\\
&+
\frac{4Mc_r^2}{|\Omega_r|\epsilon^2}
\sum_{\mathbf n\in\mathcal S_M}
\int_{\Omega_r}\int_{\Omega_Q}
\left|
\mathcal F^{-1}_{\mathbf q\to\mathbf r}
\{I(\mathbf q,\mathbf Q)\}
\right|^2
\left|
\mathcal{HG}_{\mathbf q\to\mathbf n}
\{K^{GT}(\mathbf q,\mathbf Q)\}
-
\mathcal F^{-1}_{\mathbf q\to\mathbf r}
\{K^{GT}(\mathbf q,\mathbf Q)\}
\right|^2
\,d^2Q\,d^2r .\\
&\leq  \frac{M}{\card{\Omega_r} \epsilon}
\int_{\Omega_r}\sum_{\mathbf n\in\mathcal S_M}\|\Phi_{\mathbf n}(\mathbf q)
-
c_r e^{\,i2\pi\mathbf q\cdot\mathbf r}\|_{L^\infty(\Omega_{\mbf q})}^2 d^2r 
\int_{\Omega_R}D(\mathbf R)^2\,d^2R\\
&+ \frac{4Mc_r^2}{|\Omega_r|\epsilon^2}\frac{\pi^2 \alpha^4}{\lambda^4}
\int_{\Omega_r} \sum_{\mathbf n\in\mathcal S_M}\left\|
\Phi_{\mathbf n}(\mathbf q)
-
e^{\,i2\pi\mathbf q\cdot\mathbf r}
\right\|_{L^\infty(\Omega_q)}^2  d^2r\int_{\Omega_R}D(\mathbf R)^2\,d^2R,
\end{align}
where $c_r =\frac{\card{\Omega_r}}{M}$.  We have upper bound of the total basis mismatch 
\begin{equation}
\begin{aligned} 
\mathcal E_{\mathrm{hgf}}&\leq 
\frac{M}{\epsilon \card{\Omega_r}}\int_{\Omega_r} \sum_{\mathbf n\in\mathcal S_M}\left\|
\Phi_{\mathbf n}(\mathbf q)
-
c_re^{\,i2\pi\mathbf q\cdot\mathbf r}
\right\|_{L^\infty(\Omega_q)}^2 d^2r 
\int_{\Omega_R}D(\mathbf R)^2\,d^2R 
\\ &+
\frac{
4\pi^2 M c_r^2\alpha^4
}{
\card{\Omega_r}\lambda^4\epsilon^2
}
\int_{\Omega_r}\sum_{\mathbf n\in\mathcal S_M} \left\|
\Phi_{\mathbf n}(\mathbf q)
-
e^{\,i2\pi\mathbf q\cdot\mathbf r}
\right\|_{L^\infty(\Omega_q)}^2d^2r 
\int_{\Omega_R}D(\mathbf R)^2\,d^2R .
\end{aligned}
\end{equation}
Since $
D(\mathbf R)^2
\le
D_{\max}D(\mathbf R),$
with
$
D_{\max}
=
\underset{{\mathbf R\in\Omega_R}}{\operatorname*{ess\,sup}} D(\mathbf R)\,
\text{and}\,
D_{\mathrm{tot}}
=
\int_{\Omega_R}D(\mathbf R)\,d^2R,
$ as well as the standard separable two-dimensional $L^2$-normalized HG basis $
\|\Phi_{\mathbf n}\left(\mbf q \right)\|_{L^\infty(\Omega(\mbf q))}\le\pi^{-1/2}.
$
since $|e^{\,i2\pi\mathbf q\cdot\mathbf r}|=1$ we get
$
 \left\|
\Phi_{\mathbf n}(\mathbf q)
-
c_re^{\,i2\pi\mathbf q\cdot\mathbf r}
\right\|_{L^\infty(\Omega_q)}^2
\le
\left(
\frac{1}{\sqrt{\pi}}
+
\frac{|\Omega_r|}{M}
\right)^2\,\text{and}\,
 \left\|
\Phi_{\mathbf n}(\mathbf q)
-
e^{\,i2\pi\mathbf q\cdot\mathbf r}
\right\|_{L^\infty(\Omega_q)}^2
\le
\left(
1+\frac{1}{\sqrt{\pi}}
\right)^2.
$
For fixed measurement settings, \(\alpha\), \(\lambda\), and \(|\Omega_r|\) are fixed constants. Hence, the adjustable parameters entering the bound are the number \(M\) of retained HG modes and the Wiener regularization parameter \(\epsilon\). 
\begin{equation}
{
\begin{aligned}
\mathcal E_{\mathrm{hgf}}
\le
\Bigg[
&
\frac{
\left(
M/\sqrt{\pi}+|\Omega_r|
\right)^2
}{\epsilon}
 +
\frac{
4\pi^2|\Omega_r|^2\alpha^4
}{
\lambda^4\epsilon^2
}
\left(
1+\frac{1}{\sqrt{\pi}}
\right)^2
\Bigg]
D_{\max}D_{\mathrm{tot}} = C_{\mathrm{hgf}} (M, \epsilon)
D_{\max}D_{\mathrm{tot}}.
\end{aligned}
}
\end{equation}
\end{document}